%% file: emnlp2023.tex
\pdfoutput=1

\documentclass[11pt]{article}

\usepackage[]{EMNLP2023}

\usepackage{times}
\usepackage{latexsym}
\usepackage{graphicx}
\usepackage{xspace} 

\usepackage[T1]{fontenc}

\usepackage[utf8]{inputenc}

\usepackage{microtype}
\usepackage{booktabs}
\usepackage{hyperref}

\usepackage{inconsolata}
\usepackage{graphicx}
\newcommand{\system}{{\textsc{RESIN-Editor}}}
\newcommand{\resin}{{\textsc{RESIN-11}}}

\NewDocumentCommand{\zixuan}
{ mO{} }{\textcolor{blue}{\textsuperscript{\textit{Zixuan}}\textsf{\textbf{\small[#1]}}}}
\NewDocumentCommand{\zoey}
{ mO{} }{\textcolor{orange}{\textsuperscript{\textit{Zoey}}\textsf{\textbf{\small[#1]}}}}
\NewDocumentCommand{\heng}
{ mO{} }{\textcolor{purple}{\textsuperscript{\textit{Heng}}\textsf{\textbf{\small[#1]}}}}

\title{\system: A Schema-guided Hierarchical Event Graph Visualizer and Editor}

\author{
 Khanh Duy Nguyen$^{1}$\thanks{\; Equal Contribution.}, Zixuan Zhang$^{1}$\footnotemark[1], Reece Suchocki$^{2}$, Sha Li$^{1}$,\\
  \textbf{Martha Palmer}$^{2}$\textbf{,} \textbf{Susan Brown}$^{2}$\textbf{,} \textbf{Jiawei Han}$^{1}$\textbf{,} \textbf{Heng Ji}$^{1}$ \\
  $^1$University of Illinois Urbana-Champaign \\
  $^2$University of Colorado Boulder \\
  \texttt{\{knguye71, zixuan11, shal2, hanj, hengji\}@illinois.edu} \\
  \texttt{\{reece.suchocki, susan.brown, martha.palmer\}@colorado.edu}\\
  }

\begin{document}
\maketitle


\input{abstract}
\input{introduction}

\input{system_overview}
\input{interface}
\input{experiment}
\input{related_work}
\input{case_study}
\input{conclusion}

\section*{Acknowledgements}
This research is based on work supported by U.S. DARPA KAIROS Program No. FA8750-19-2-1004. The views and conclusions contained herein are those of the authors and should not be interpreted as necessarily representing the official policies, either expressed or implied, of DARPA, or the U.S. Government. The U.S. Government is authorized to reproduce and distribute reprints for governmental purposes notwithstanding any copyright annotation therein.

\bibliography{anthology,custom}
\bibliographystyle{acl_natbib}




\end{document}

%% file: abstract.tex
\begin{abstract}
In this paper, we present \textsc{RESIN-Editor}, an interactive event graph visualizer and editor designed for analyzing complex events.
Our \system~system allows users to render and freely edit hierarchical event graphs extracted from multimedia and multi-document news clusters with guidance from human-curated event schemas.
\system's unique features include hierarchical graph visualization, comprehensive source tracing, and interactive user editing, which is more powerful and versatile than existing Information Extraction (IE) visualization tools.
In our evaluation of \system, we demonstrate ways in which our tool is effective in understanding complex events and enhancing system performance.
The source code\footnote{Data and source code: \url{https://github.com/blender-nlp/RESIN-Editor}}, a video demonstration\footnote{Video demonstration: \url{https://www.youtube.com/watch?v=fmW-GwPMrw0}}, and a live website\footnote{Live website: \url{https://blender-nlp.github.io/RESIN-Editor/}} for \system~have been made publicly available.
\end{abstract}

%% file: introduction.tex
\section{Introduction}


Complex events, such as an international negotiation or a disease outbreak, can take place over a prolonged period of time spanning from weeks to even months. 
Typically, such complex events can be further deconstructed into many atomic sub-events, where each occurs at a specific time and place. 
Fully modeling such complex events requires an understanding of the temporal, logical, and hierarchical connections among many sub-events, making the task of event modeling difficult for existing Information Extraction (IE) visualization and analysis tools.
Recent research efforts~\cite{wang-etal-2020-joint,du-etal-2022-resin, prediction} extend the event understanding paradigm 
to model event-event relations and 
propose to use event schemas to guide the organization of complex event structures and predict possible future events. 
However, convenient visualization and editing tools specifically designed to handle such types of complex events are still largely undeveloped.

A visualization and editing tool is essential for developing and improving an IE system.
By utilizing such a tool, system developers can conduct a qualitative error analysis, which involves human analysts carefully investigating any missing or spurious errors, tracing these inaccuracies back to the original documents, and record any proposed modifications or suggestions. 
Existing tools designed for atomic events~\cite{openievisual,li-etal-2019-multilingual} are limited when directly applied to complex events, which are much more intricate with complicated event-event relations and multi-document multi-media source inputs.
Furthermore, prior analysis tools lack support for interactive user editing, which poses a challenge for model developers to comprehend any real feedback from actual users.

To address the previously discussed challenges and accelerate research of complex events, we present the \system, an analysis interface that allows users to visualize, analyze, and freely edit the multi-granularity graphical event narratives extracted from multi-document multimedia news clusters.
Compared with existing analysis interfaces, our tool has the following three significant advantages:
1) \textbf{\emph{Hierarchical Graph Visualization}}.
Our tool is able to generate a comprehensive visualization of the extracted hierarchical event graph (as shown in Figures \ref{fig:overview} and~\ref{fig:interface}).
This graph not only includes atomic events and entities, but also includes all event-event hierarchical, temporal, and logical relations of the entire complex event.
Such an informative visualization can offer human analysts a rapid and comprehensive understanding of any complex event, significantly improving the efficiency of our error analysis process.
Moreover, our graphical visualization is initialized with carefully-curated schemas, and the extracted events are grounded to the schema whenever possible, which improves the reliability of our error analysis.
2)~\textbf{\emph{Comprehensive~Source Tracing}}.
Our tool is able to show multimedia provenances for all extracted events, entities, and relations, including source documents, text spans, images, and bounding boxes. 
These trace-back provenances are presented in a clear-structured and tabular format, which greatly simplifies the users' task of understanding and evaluating the performance of the extraction system.
3) \textbf{\emph{Interactive User Editing}}.
Most importantly, with our analysis tool, users are allowed to edit and manipulate almost every extracted element depicted in the visualized event graph and save their edited event graph for downstream usage. 
This includes not only the node attributes of events and entities, such as names, types, and description sentences but also the graphical structures, like the entity-entity relations and event-event temporal links.
Through open and direct interactions with intended users, our tool equips model developers with a comprehensive and precise understanding of actual user requirements. \\

Our main contributions can be summarized as: 
\begin{itemize}
    \item A novel development to a multimedia analysis tool for complex event understanding, which enables multi-granularity visualization, comprehensive source document tracing, and interactive user editing.
    
    \item Extensive empirical studies show that our tool is significantly more effective than existing tools in understanding and analyzing complex events.

\end{itemize}

%% file: system_overview.tex
\begin{figure*}
  \includegraphics[width=0.98\textwidth]{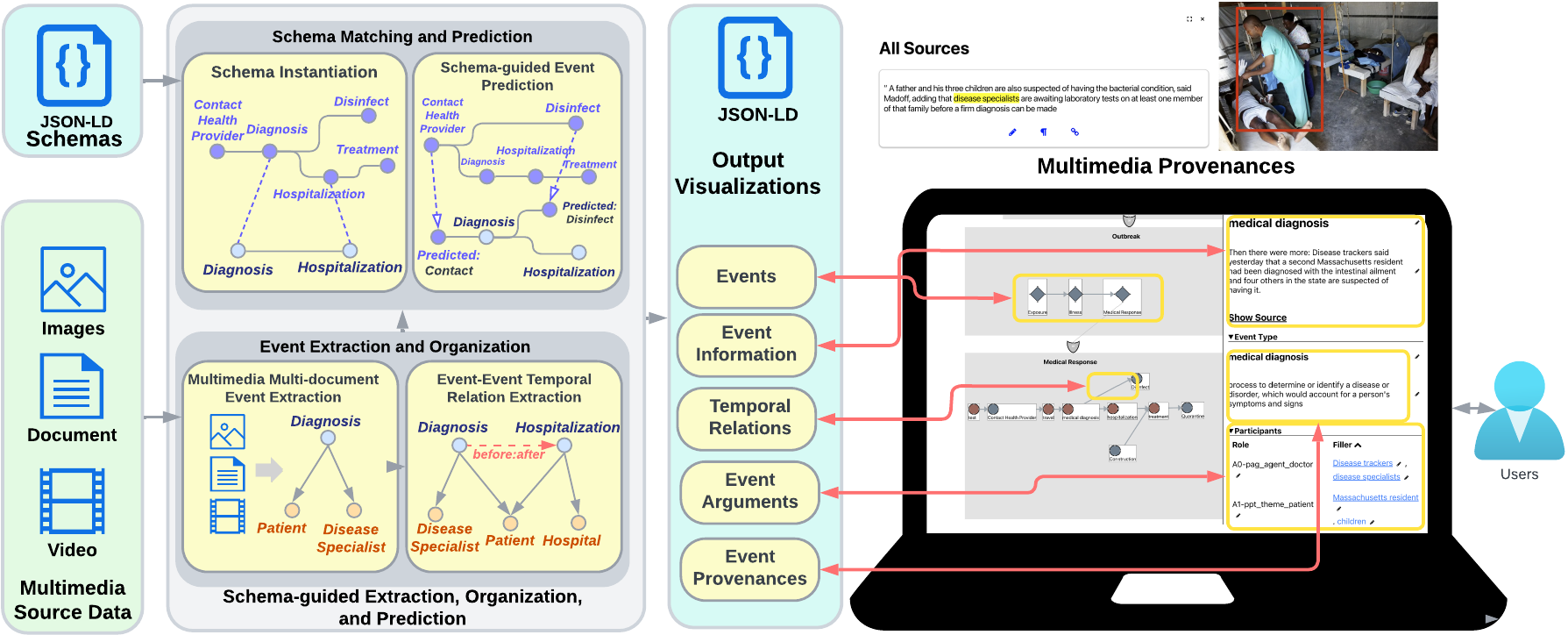}
  \caption{Overview of our \system \xspace system. Given a multimedia multi-document news cluster and the schema graph, our system extracts the atomic events, organizes them into an instance event graph, performs schema-guided matching and prediction, and finally interactively visualizes the results for user editing.}
  \label{fig:overview}
\end{figure*}


\section{Backend System Overview}
The \system \xspace system 
provides an interface for event graph visualization with the added capability of graph editing. 
The visualized graph is an instantiated schema graph, which results from matching an extracted event graph with an event schema. An event schema is a prototypical description of the usual progression of a particular type of complex event, which can be induced from historical news documents.

In this section, we present an overview of how this event graph is generated, with the main components being multimedia event extraction, cross-document event organization, and event prediction in hierarchical schemas.

\paragraph{Multimedia Event Extraction}

The input to our system is a multimedia news cluster. This cluster includes multiple news documents with related images from a variety of news sources, all describing a single news event.
From this cluster, we first perform event extraction for atomic events. Following the convention of ACE-2005\footnote{\url{https://www.ldc.upenn.edu/collaborations/past-projects/ace}}, each event is composed of an event trigger as well as several event arguments. 
Since we would like our system to work for all news scenarios, we employ GLEN~\cite{zhan2023glen}, the event trigger extraction system that works for 3k+ event types, making it the most comprehensive event detection system to date. 
We then extract event arguments with a template-based generation approach~\citep{li-etal-2021-document} distilled from a large language model that is prompted to extract arguments as code structures~\cite{wang-etal-2023-code4struct}. To handle visual input, we use \citep{li-etal-2020-cross} to extract events from images and merge them with events extracted from text. 
We represent the atomic events from the multimedia event extraction system as star-shaped graphs such as the \emph{Diagnosis} event shown in Figure~\ref{fig:overview}.

\paragraph{Cross-Document Event Organization}
After extracting atomic events and entities, all event and entity mentions that correspond to the same real-life event and entity are merged through coreference resolution~\cite{lai-etal-2021-context}, and linked to Wikidata through entity linking~\cite{lai-etal-2022-improving}
We then conduct event-event temporal relation extraction~\cite{jin-etal-2023-toward} to organize the events into a temporal event graph.
We refer to the resulting event graph as the instance event graph, which serves as a semantic representation of the input news cluster in aggregate. 
The event graph exemplifies unfolding narratives within the given news cluster, as well as connects entities across articles.

\paragraph{Matching and Prediction with Hierarchical Schemas}
After extracting the instance event graph, we perform event instantiation and prediction with the help of event schemas.
The schema is typically depicted as a graph in which the nodes are generalized event types linked by inter-event relations, and our task is to instantiate the schema with our instance event graph and make predictions for those unmatched events.
As compared to previous systems~\cite{du-etal-2022-resin, prediction}, we now facilitate event matching and prediction on hierarchical schema structures, where the events in the schema are presented with parent-child relations to represent the different granularity of events.
All of the extracted event nodes in our system are grounded in the structured event hierarchy, either by matching to a schema node, or being attached as a child event to a schema node. Event and entity nodes in our extracted event graph are matched top-down, enabling event prediction based on our graphical neural network event prediction system~\cite{prediction}.

%% file: interface.tex
\section{Graphical Interface}

\begin{figure*}
    \centering
    \includegraphics[width=.8\textwidth]{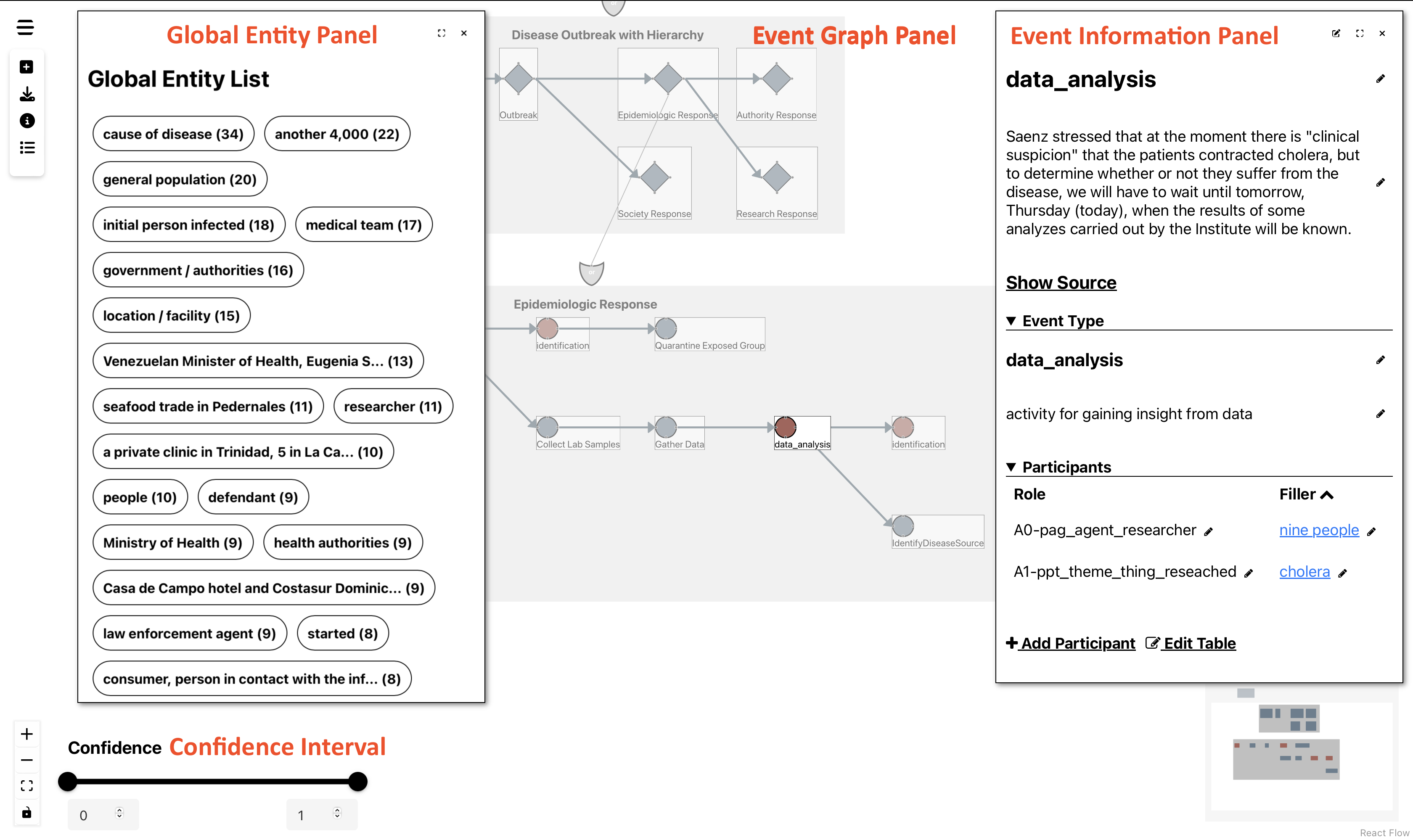}
    \caption{The \system \xspace interface with \emph{Cholera Decease Outbreak} event graph. 
    }
    \label{fig:interface}
\end{figure*}

\begin{figure}
    \centering
    \includegraphics[width=0.45\textwidth]{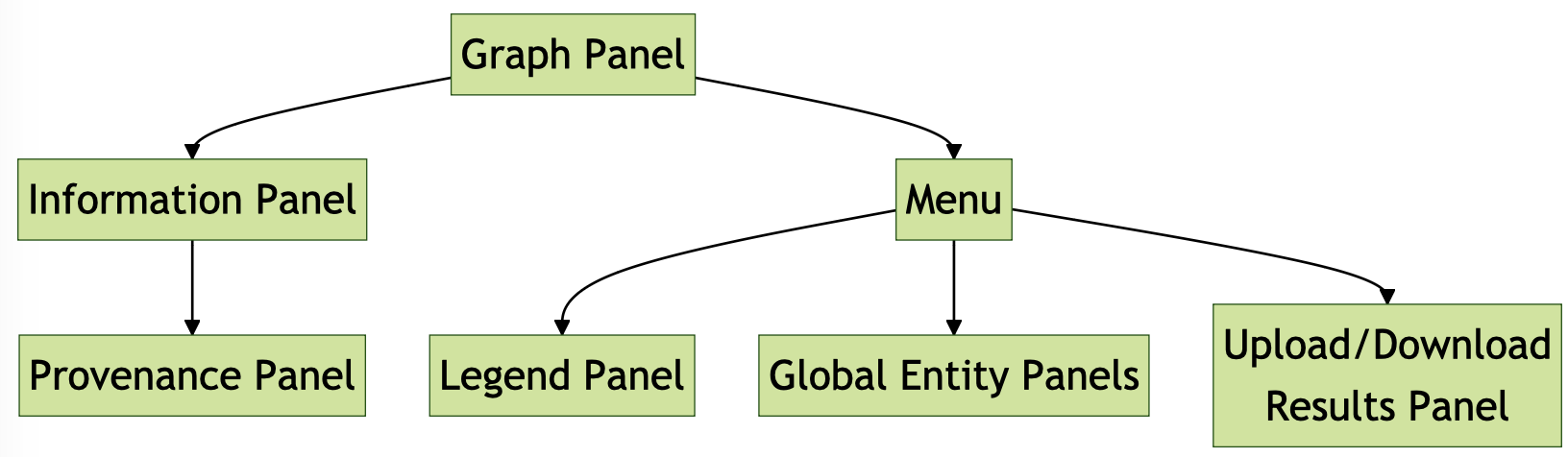}
    \caption{The View Panel Hierarchy illustrates the call sequence of the panels. The most significant panel is located at the top, and subsequent panels are activated from their immediate parent panels. }
    \label{fig:panel}
\end{figure}


Our graphical interface visualizes the instantiated schema graph after matching and prediction along with the source documents that provided provenance for the extracted events. 
The interface (Figure \ref{fig:interface}) is organized in the panel hierarchy illustrated in Figure \ref{fig:panel}.
The graph panel, which is the canvas for showing the event graph, is the primary access point.
The overlay panels, such as the entity table, provenance, and mini-map, enhance the exploration experience with contextually relevant information or provide navigation support.

\subsection{Interactive Graph Panel}
We adopt a graph-based representation for the schema-matching results to provide a streamlined and efficient error analysis interface for event analysts. 
For each event, we use color coding to illustrate whether the event is extracted from the source document (source-only), from the schema (predicted), or found in both (matched). 

\begin{figure}
    \centering
    \includegraphics[width=0.3\textwidth]{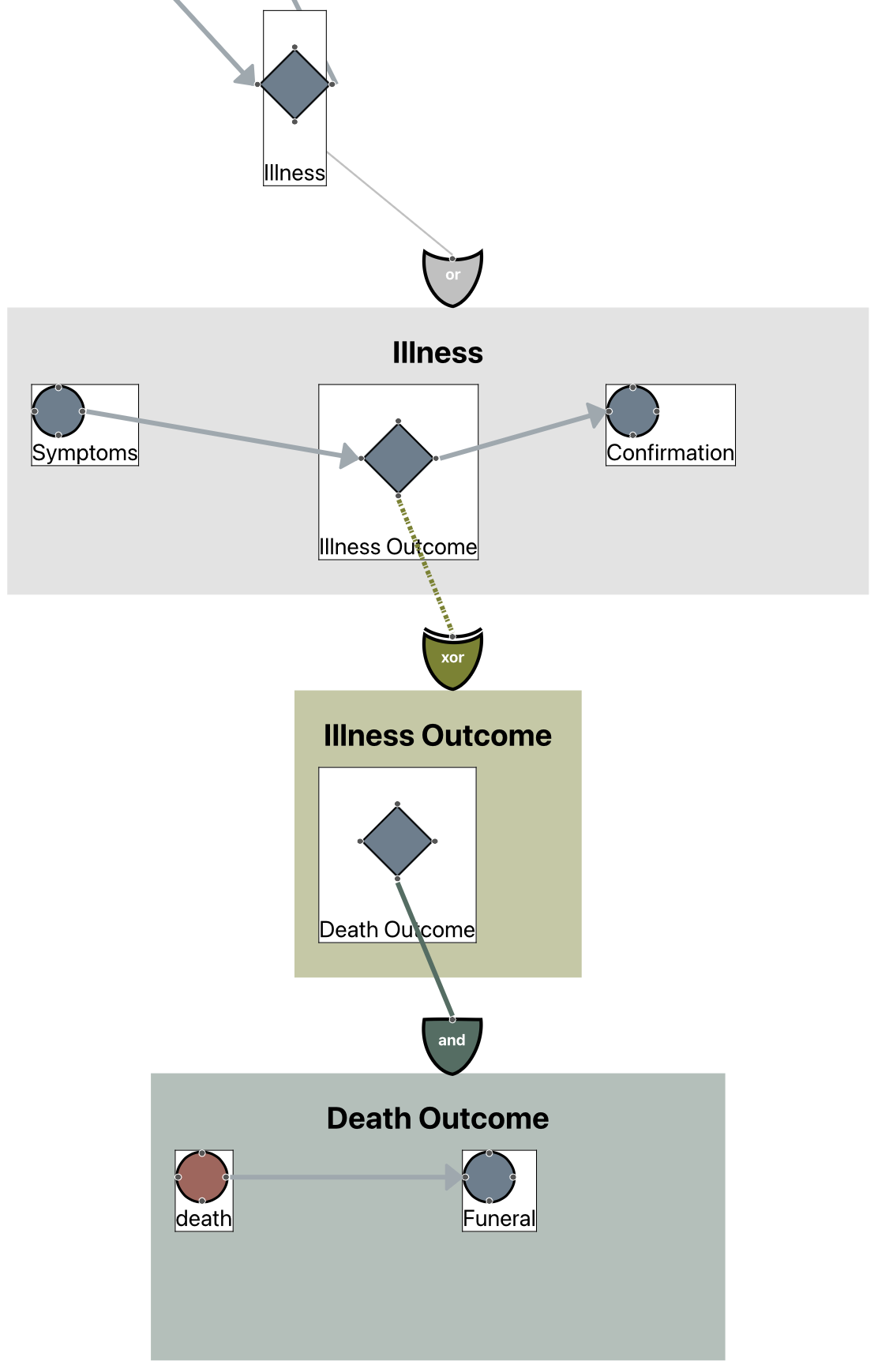}
    \caption{After an \emph{Illness} event, possible subsequent events include \emph{Symptoms}, \emph{Illness Outcomes}, or \emph{Confirmation }. Following the \emph{Illness Outcomes} event, only a single \emph{Death Outcomes} event can occur. The \emph{Death Outcomes} event then invariably leads to both \emph{Death} and \emph{Funeral} events.}
    \label{fig:gates}
    \vspace{-0.3cm}
\end{figure}

To assist analysts or system developers in making informed decisions, we find it critical to present the relations between events clearly and disentangle different types of relations. 
We support 3 types of relations among events in our visualization tool:
\begin{itemize}
    \item \textbf{Hierarchical event-event relations.} Parent-child relations between events are visualized by arranging them on the vertical axis: parent events are shown above their children events. Primitive events, which are leaf nodes in the graph, are displayed as circular nodes, whereas parent events are displayed as diamond-shaped nodes. Initially, we only show the top-level events as an overview of the scenario. Users can click on the parent event to expand the graph and dive deeper into the details of the event. 
    \item \textbf{Temporal relations.} 
    We order events with temporal relations along the horizontal axis from left to right. In the case of misaligned temporal relations, the user can change the event order by editing the temporal edge. 
    \item \textbf{Logical relations. } 
    We support the usage of logical relations defined on a group of events to enforce stronger logical constraints. The \emph{OR gate} signifies that one or more events in the group can occur, the \emph{AND gate} necessitates the occurrence of all events, and the \emph{XOR gate} permits exclusively one event to happen.
     Given the common use of digital logic gates in computer science and the conceptual analogy they provide, we employ them to depict logical gates in our event graph (Figure \ref{fig:gates}). 
    
\end{itemize}

To improve navigation efficiency in the Graph Panel, we include
direct navigation through drag panels and event dragging,
a zoom function for detailed exploration, and
a minimap for a macroscopic overview of the graph.
More detailed information and context can be accessed by clicking on an event node, which triggers an information panel.


\subsection{Information Panel}
The individual observer without assistance has significant limitations regarding how much information they can perceive, process, and retain \cite{Miller1956}.
To avoid interface clutter, we partition the event's information into three distinct sections:

\begin{figure}
    \centering
    \includegraphics[width=0.47\textwidth]{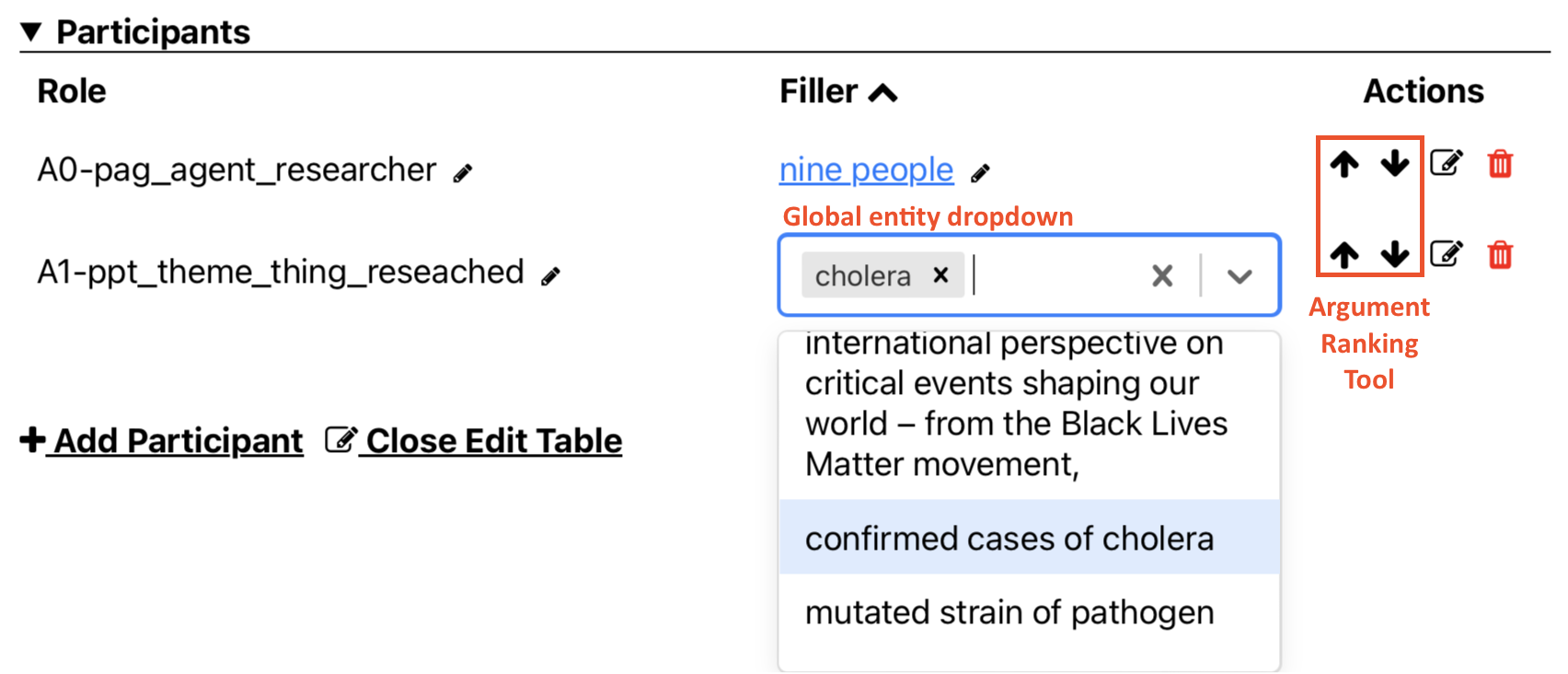}
    \caption{Updating the argument lists of the \emph{Data Analysis} event, incorporating the \emph{Confirmed Cases of Cholera} entity into the theme research.}
    \label{fig:participant-list}
\end{figure}

\paragraph{Event Description}
Include the event name and the description of the event. Furthermore, descriptions of events can be viewed swiftly by hovering over the events on the Graph Panel.
\paragraph{Event Type}
The event type information is either from the event schema or the event detection system. We use the XPO ontology \cite{spaulding2023darpa} for the scope of event types covered by our system. 
This information might prove helpful in cases where the event description is insufficiently descriptive. 
\paragraph{Argument Table}
Each event is associated with multiple argument roles as defined by the ontology, and each argument role can have multiple fillers. 
We present this information in the  argument table which consists of rows of arguments (e.g., location, places, participants) and their associated entities, as shown in Figure \ref{fig:participant-list}. 
We include only the entities backed by sources, leaving out generic ones. In editing mode, the user can reorder the arguments per their needs, ensuring that the most relevant arguments are prioritized at the top. Furthermore, the edit mode provides a comprehensive list of entities from which analysts can add any relevant missing entities as arguments.
Each information display includes an editing function, allowing analysts to adjust the presented event information while analyzing schema-matching results. 

\subsection{Provenance Panel}

\begin{figure}
    \centering
    \includegraphics[width=0.48\textwidth]{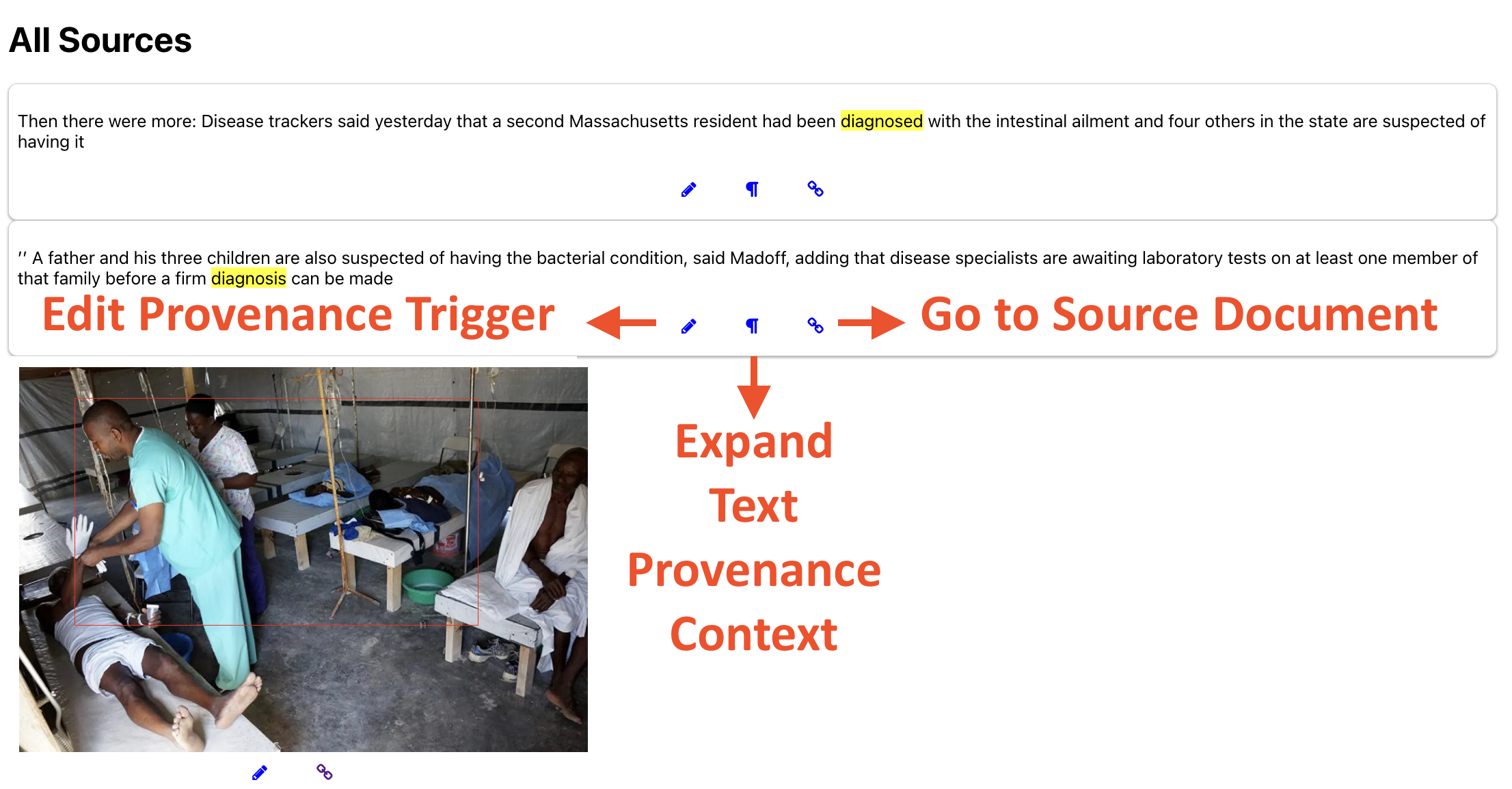}
    \caption{Multimedia provenances support \emph{Disease Specialist} entity.}
    \label{fig:provenance}
\end{figure}

Identifying the source document that validates schema matching is essential for analysts during error analysis because it helps to verify any inaccuracies either from the IE system or in matching process. Hence, in the provenance panel (Figure \ref{fig:provenance}), we display the text and image provenances for source-only events, matched events, and entities. 
We utilize text highlighting within the paragraph to indicate text provenances, allowing the user to find the relevant citation in the source document easily. Additionally, image provenances employ a bounding box to indicate the referenced entity in the image.
Under each piece of provenance, we provide three functions: 
\paragraph{Edit the provenance mentions.}
Analysts can substitute inaccurate text spans in the source document with the correct trigger span for text provenance. For image provenance, the bounding box can be modified to suit the entity in the image better. For instance, in Figure \ref{fig:provenance}, analysts can realign and resize the bounding box to correspond with the nurse image. This adjustment allows for adequately presenting the \emph{Disease Specialist} entity inside the provenance of the \emph{Diagnosis} event.

\paragraph{Expand the provenance context.}
Analysts can expand and view the entire paragraph containing the extracted provenance, but only for text provenance.
\paragraph{Go to source document.} 
The feature enables tracing back to the original documents containing the sentence, providing analysts with additional contextual information about the provenance. This tool becomes essential when analysts need to authenticate the provenance's integrity.

\subsection{Event Filtering Tools}
To ease navigation through numerous graph events, we develop these two event filters:

\paragraph{Global Entity List}
This feature aids analysts in isolating entity-related events by reducing the opacity of unrelated events. For instance, selecting \emph{Cholera} entity in Figure \ref{fig:interface} emphasizes the connected \emph{Data Analysis} event, while other events become dim. 
Moreover, the global entities list ranks entities based on their occurrence in graph events, highlighting those most frequently mentioned across all events.

\paragraph{Confidence Interval}

The schema-matching process between the instance graph and the schema assigns a confidence value to each event from its schema matching and prediction component~\cite{prediction}. 
This tool is beneficial when analysts aim to discover matching events with less certainty. By selecting a lower confidence interval, analysts can monitor all events within that range and investigate those with low confidence values.

%% file: experiment.tex
\section{Experiments}
We present our empirical experiments with multiple document clusters to evaluate the effectiveness of our editor.
\paragraph{Evaluation Setup}
We use two high-impact news scenarios \emph{Disease Outbreak} and \emph{Terrorist Attacks} with four newsworthy complex events, where the detailed statistics are shown in Table~\ref{tab:data}.

\paragraph{Results}
Based on the results shown in Table~\ref{table:results}, \system~ significantly outperforms the traditional method of direct source document analysis, showing considerable improvements across all main IE tasks in our backend system. 
Such an improvement is particularly notable in the prediction task, probably because the original documents offer limited supporting information for predicting new events. 
However, our visualizer can provide a well-organized graphical event timeline, which is greatly beneficial for comprehending the events and making accurate predictions.

\begin{table}[htbp]
	\centering
    \footnotesize
	\begin{tabular}{ccc}
		\toprule[1pt] 
		  News Cluster & \# Docs & \# Images \\
		\midrule[1pt]
            Cholera outbreak, Dominica & 13 & 114\\
            E. coli outbreak, USA & 16 & 118\\
            Legionnaires outbreak, USA & 14 & 66\\
            Mogadishu bombings, Somalia & 13 & 57\\
		\midrule[1pt]

	\end{tabular}
	\normalsize
	\caption{Statistics of the news document clusters.}
 \vspace{-0.4cm}
	\label{tab:data}
\end{table}

\begin{table}[htbp]
	\centering
    \small
	\begin{tabular}{c|ccc}
		\toprule[1pt] 
		  Tasks & Prec & Rec & F1  \\
		\midrule[1pt]
            Event Triggers (documents)& 75.5 & 69.0 & 72.1\\
            Event Triggers (visualizer) & 84.9 & 76.3 & \textbf{80.4}\\
		\midrule[0.5pt]
            Event Arguments (documents)& 68.0 & 58.6 & 63.0\\
            Event Arguments (visualizer) & 84.0 & 72.4 & \textbf{77.8}\\
		\midrule[0.5pt]
            Schema Matching (documents)& 67.9 & 62.3 & 65.0\\
            Schema Matching (visualizer) & 71.1 & 64.5 & \textbf{67.6}\\
		\midrule[0.5pt]
            Prediction (documents)& 62.9 & 50.0 & 55.7\\
            Prediction (visualizer) & 93.1 & 66.7 & \textbf{77.8}\\

		\midrule[1pt]

	\end{tabular}
	\normalsize

	\caption{The improved IE task performances using our designed visualizer compared with directly looking into the source documents. We focus on the four main IE tasks in our backend system, and the performances are characterized in precision, recall, and F1 scores.}
        \vspace{-0.4cm}
	\label{table:results}
\end{table}

%% file: related_work.tex
\section{Related Work}


In the IE visualization domain, \cite{jenkins-etal-2023-massively, vacareanu-etal-2022-human} offer event-focused indexed documents, while others \cite{ma-etal-2021-eventplus, ning-etal-2018-cogcomptime} use temporal-relation entity graphs. However, both methods lack hierarchical event organization which leads to clutter in the graph display. Additionally, despite presenting entity provenance triggers, they are not able to assist in amending extraction errors.

In terms of schema visualization, \citet{mishra-etal-2021-graphical} introduce a graph-based interface tailored to depict schemas, emphasizing event order and entity relations.  While the interface allows users to edit schemas, the user must come up with editing suggestions based on his/her own knowledge without any help from the interface. 
More recently, \citet{zhang-etal-2023-human} introduce a human-in-the-loop interface that employs pre-trained LLMs to help create schema graphs detailing event sequences in scenarios. Whereas the use of LLMs can alleviate the manual effort of the user, the LLM-generated schema is not grounded in real news documents and may be prone to hallucination. 

Existing tools mainly focus on either IE extraction or schema visualization. Inspired by the \resin\ system \cite{du-etal-2022-resin}, which generates schema-guided event graphs from multi-modal news documents, we introduce a unified tool that integrates both the schema and the extracted event graph, providing a comprehensive visualization for complex event analysis. By matching the extracted event graph with the human-curated schema, we are able to effectively organize the complex event and guide users in providing informed feedback on both the IE results and the schema. 

%% file: conclusion.tex
\section{Conclusion and Future Work}


To the best of our knowledge, \system~is the first graphical user interface for editing and visualizing schema-guided event graphs.
Our tool assists users by highlighting the matching results, allowing them to fix discrepancies between event attributes and their argument entities and resolve disparities between extracted entities and their multimedia provenance. 
Our results show that our solution outperforms the traditional method of error straight from their sources in primary IE tasks. While we recognize the importance of comparing with other tools, our current experimentation is limited due to time constraints. We plan to conduct extensive experiments in the future to validate our results further.

In future work, we aim to develop a ranking system that considers the relevance of multimedia provenance so that users can quickly view the most related sentence/image. We also plan to integrate the event grounding tool, so that users can easily add new events by providing natural language descriptions, and the system will provide event type suggestions and fill in the argument template. 
